\documentclass[12pt,a4paper]{article}  
\usepackage[dvips]{graphicx}
\begin{document}
\begin{titlepage} 
\begin{flushright} 
UAB--FT--524\\ 
hep-ph/0204338\\
October 2002 
\end{flushright}
\vspace*{1.6cm} 
 
\begin{center} 
{\Large\bf  
On the mass, width and coupling constants of the $f_0(980)$}\\ 
\vspace*{0.8cm} 
 
R.~Escribano$^1$, A.~Gallegos$^2$, J.L.~Lucio M.$^2$, G.~Moreno$^2$ and 
J. Pestieau$^3$\\ 
\vspace*{0.2cm} 
 
{\footnotesize\it 
$^1$Grup de F{\'\i}sica Te{\`o}rica and IFAE, Universitat Aut{\`o}noma de Barcelona,\\
E-08193 Bellaterra (Barcelona), Spain\\ 
$^2$Instituto de F{\'\i}sica, Universidad de Guanajuato,\\ 
Lomas del Bosque \#103, Lomas del Campestre, 37150 Le{\'o}n, Guanajuato, Mexico\\
$^3$Institut de Physique Th{\'e}orique, Universit{\'e} Catholique de Louvain,\\ 
Chemin du Cyclotron 2, B-1348 Louvain-la-Neuve, Belgium} 
\end{center} 
\vspace*{1.0cm} 
 
\begin{abstract}
\noindent
Using the pole approach we determine the mass and width of the $f_0(980)$,
in particular we analyze the possibility that two nearby poles are associated to it.
We restrict our analysis to a neighborhood of the resonance,
using $\pi\pi$ data for the phase shift and inelasticity, and the invariant mass
spectrum of the $J/\psi\rightarrow\phi\pi\pi, \phi K\bar K$ decays.
The formalism we use is based on unitarity and a generalized version of the 
Breit-Wigner parameterization.
We find that a single pole describes the $f_0(980)$, the precise position
depending upon the $\pi\pi$ data used (see Eq.~(\ref{poles}) of the main text).
As a byproduct, values for the $g_{f_0\pi\pi}$ and $g_{f_0K\bar K}$
coupling constants are obtained.
\end{abstract} 
\end{titlepage} 

\section{Introduction} 
According to the Review of Particles Physics (RPP) \cite{Groom:2000in},
the mass of the $f_0(980)$ scalar resonance is $980\pm 10$ MeV
whereas the width ranges from 40 to 100 MeV.
The reasons for such an uncertainty in the width are the great amount and variety of experimental data
and the different approaches used to extract the intrinsic properties of the resonance.
To these points we could also add the lack of a precise definition of what is meant by mass and width,
although there seems to be consensus in using the pole approach,
where the mass and width of the resonance are found from the position of the nearest pole 
in the $T$-matrix (or equivalently, the $S$-matrix).
However, even if this approach is adopted the final results differ on the number,
location and physical interpretation of the poles.
This is because the pole approach is not enough to completely fix the framework
needed to perform the resonance analysis; in fact there are different formalisms that made use of it. 
An example is field theory, where a finite imaginary part of the propagator arises after 
Dyson summation of the one-particle-irreducible diagrams contributing to the two-point function.
Unitarity also implies a general complex structure of the $T$-matrix
in terms of which the pole approach becomes relevant to define the mass and width of the resonance.
However, the general solution to the unitarity constraint has no implications regarding 
the number and/or locations of the poles.
Thus, most analyses using the pole approach must involve further assumptions.

Far from physical thresholds, the identification of the mass and width of a resonance
in terms of the nearest pole in the $T$-matrix is not ambiguous.
However, when the resonance lies in the vicinity of a threshold this identification
is not so obvious and more than one single pole can be required for a correct description
of the resonance (see for example Ref.~\cite{Bhattacharya:1993gr}).
This could be the case for the $f_0(980)$ whose mass is very close to the $K\bar K$ threshold.
In Ref.~\cite{Morgan:1993td}, Morgan and Pennington (MP) use a formalism general enough
to avoid any assumption about the number of poles associated to a resonance.
For the particular case of the $f_0(980)$, their exhaustive analysis leads to the conclusion 
that the $f_0(980)$ is most probably a Breit-Wigner-like resonance 
---with a narrow width $\Gamma\sim 52$ MeV---
which can be described in terms of two nearby poles (in the second and third sheets).
Moreover, the precision data coming from $J/\psi\rightarrow\phi (MM)$ decays, 
where $(MM)$ stands for $\pi\pi$ or $K\bar K$,
play an essential r\^ole as a crucial check in favour of the two-pole description 
of the $f_0 (980)$ and disfavour the cases with one or three poles.
These results indicate that the description of the $f_0(980)$ using a Breit-Wigner parametrization
seems to be appropriate and should give results similar to those in
Refs.~\cite{Morgan:1993td,Morgan:1993rn}.  
However, an existing analysis using a Breit-Wigner parametrization performed by
Zou and Bugg (ZB) \cite{Zou:1993az} concludes that the $f_0(980)$ is most likely a resonance
with a large decay width ($\sim 400$ MeV) and a narrow peak width ($\sim 47$ MeV).
Later on, MP \cite{Morgan:1993rn} and ZB \cite{Zou:1994ea} have both confirmed their former results,
leaving the agreement between the two approaches as an open question.

The pole approach formalism has been successfully applied to hadronic resonances
such as the $\rho(770)$ \cite{Bernicha:1994re,Benayoun:1998ex,Feuillat:2001ch} and 
the $\Delta (1232)$ \cite{Bernicha:1996gg}.
An important advantage of this formalism is that it yields process and background independent results.
This background independence is only valid if other resonances are not present within 
the kinematical region under consideration\footnote{
The amplitude associated to  a given resonance is not expected to describe the physics
in a large kinematical region (compared to the width of the resonance) where additional
resonances can exist.}.
Thus, if one insists on this point, as we will, it is important to restrain the analysis 
to a neighborhood of the resonance under study. 
Due to the previous consideration, we exclude from our analysis of the $f_0(980)$ the
central production data $pp\rightarrow p(\pi\pi,K\bar K)p$ \cite{Barberis:1999cq}--\cite{Alde:1997ri},
which covers a much wider energy range and whose phenomenological description requires not only 
$\pi\pi$ and $K\bar K$ scattering but also a production mechanism involving many parameters.
We also exclude the experimental data on $J/\psi\rightarrow\omega (\pi\pi,K\bar K)$ decays
because the $f_{0}$ signal is too weak \cite{Augustin:1988ja,Falvard:1988fc}\footnote{
After submission of this manuscript the BES Collaboration has confirmed these results
\protect\cite{Wu:2001vz}.}
and the inclusion of these data in the fit would require further parameters\footnote{
A phenomenological analysis of the $J/\psi\rightarrow VS$ decays assuming $U(3)$ symmetry predicts
that for ideally mixed scalar and vector states the values of the relevant coupling constants are:
$g_{J/\psi\phi f_{0}}=g_{J/\psi\omega\sigma}=1$ and $g_{J/\psi\omega f_{0}}=g_{J/\psi\phi\sigma}=0$.
The small departure from ideal mixing of the $\phi$ and the $f_{0}$ states explains,
in a first approximation, the importance of the $f_{0}$ contribution to the
$J/\psi\rightarrow\phi (\pi\pi,K\bar K)$ decays 
and, on the contrary, the minor r\^ole played in the $J/\psi\rightarrow\omega (\pi\pi,K\bar K)$ decays
\protect\cite{Meissner:2000bc}.}.

Our purpose in this work is to extract the mass and width of the $f_0(980)$ scalar resonance
using the pole approach.
As a byproduct, we also obtain values for the coupling constants 
$g_{f_0\pi\pi}$ and $g_{f_0 K\bar K}$.
We suggest that, for narrow resonances as the $f_0(980)$, the complete one-loop scalar propagator
must be used in the pole equation.
Thus, our analysis is based on a generalized Breit-Wigner description of a scalar resonance
coupled to two channels,  not only satisfying unitarity but also including loop effects.
As far as the pole is concerned, we pay special attention to the possibility of describing the
$f_0(980)$ in terms of more than one pole,
a phenomenon which is known to occur when the mass of the resonance is close to a threshold.
Nevertheless, it is worth remarking that the need of two poles is not guaranteed,
it strongly depends upon the precise value of the renormalized mass of the resonance and 
its coupling constants to the two channels.

Values of the renormalized mass $m_R$ and the coupling constants $g_{f_0\pi\pi}$ and $g_{f_0 K\bar K}$ 
are obtained from a fit to experimental data including the $\pi\pi$ phase shift and inelasticity as 
well as the $J/\psi\rightarrow\phi\pi\pi(K\bar K)$ invariant mass spectra. We then look for poles in 
the four Riemann sheets associated to a resonance coupled to two channels. Our conclusion is that the 
$f_0(980)$ can be described in terms of a single pole whose precise position depends upon the $\pi\pi$
data used (see Eq.~(\ref{poles}) of the main text for details).

\section{Formalism}
Before discussing the formalism used for the particular case of the $f_0(980)$,
it is convenient to briefly summarize two well-known definitions of mass and width 
of a given resonance, both widely used in the hadron physics literature
(see Ref.~\cite{Bernicha:1996gg} and references therein). 
One definition, known as the \emph{conventional approach}, is based on the behavior of the 
phase shift of the resonance as a function of the energy,
while the other, known as the \emph{pole approach}, is based on the pole position of the resonance,
which as discussed in the introduction includes several approaches. 
We will not consider here the powerful formalism developed in Ref.~\cite{Morgan:1993td}
since it goes beyond a Breit-Wigner-like description of the resonance, to which our analysis is
restricted.
In this sense, it is worth noticing  that these more powerful methods provide further support
to the Breit-Wigner description of the $f_0(980)$.
Thus, the analysis carried in this paper is more restricted in scope, although it turns out to be 
general enough for the $f_0 (980)$ case.

In the conventional approach, the mass and width of the resonance are defined 
in terms of the phase shift $\delta$ 
as\footnote{We use the subindexes $\delta$ and $p$ to denote the mass and width 
of the resonance in the conventional and pole approaches respectively.}
\begin{equation}
\delta(s=M_\delta^2)=90^\circ\ ,\quad 
\Gamma_\delta=\frac{1}{M_\delta}\left[\frac{d\delta(s)}{ds}\right]_{s=M_\delta^2}^{-1}\ ,
\end{equation}
respectively.
Since the phase shift is extracted from direct comparison with experimental data,
the decay width defined in this way is usually called the 
\emph{visible} or \emph{peak width}.
For an elastic Breit-Wigner (BW) resonance the phase shift is chosen 
as\footnote{A $s$-dependent width $\Gamma_\delta(s)$ in
Eq.~(\protect\ref{tandelta}) is mandatory when a background around the resonance
is assumed \protect\cite{Bernicha:1996gg}.}
\begin{equation}
\label{tandelta}
\tan\delta(s)=-\frac{M_\delta \Gamma_\delta(s)}{s-M_\delta^2}\ ,
\end{equation}
which leads to the partial-wave amplitude
\begin{equation}
a=\frac{e^{2i\delta}-1}{2i}=
-\frac{M_\delta \Gamma_\delta(s)}{s-M_\delta^2+i M_\delta \Gamma_\delta(s)}\ ,
\end{equation}
where $s$ is the center-of-mass energy squared.

In the pole approach (or $S$-matrix approach), the resonance shows up as a pole in the amplitude
\begin{equation}
\label{apole}
a=\frac{R}{s-s_p}+B\ ,
\end{equation}
where the two terms correspond to the resonant and background contributions
separated according to Refs.~\cite{eden,Stuart:1991xk}.
Eq.~(\ref{apole}) is understood as a power series expansion of the amplitude around $s_p$,
therefore, in order this description to make sense, the background around the pole
(which is fixed from the fit to experimental data) should be a smooth function of $s$
affecting minimally the pole position.
In this approach, the mass and width of the resonance are defined in terms of the 
pole position $s_p$
as\footnote{The relationship between the mass and width parameters defined in the
conventional and pole approaches can be found in
Ref.~\protect\cite{Bernicha:1996gg}.}
\begin{equation}
s_p=m_p^2-i m_p\Gamma_p\ .
\end{equation}
The pole approach provides a definition for the parameters of an unstable particle which is
\emph{process independent} (independent of the process used to extract them) and also
\emph{background independent} 
(different parametrizations of the background will hardly modify the values obtained
for the pole parameters of the resonance).

In the remaining of this section we pursue the pole approach for the case of a 
resonance coupled to two channels, including furthermore the possibility of a
strongly $s$-dependent width due to the opening of a second two-body threshold.
Two ingredients are required in order to build the scattering amplitude to be used
in our formalism: unitarity and the complete one-loop scalar propagator. 

Concerning the first of the ingredients, unitarity sets stringent constraints
on the amplitudes needed for the description of a resonance coupled to several
channels.
The correct incorporation of these constraints into the $S$-matrix is compulsory
for an adequate analysis of experimental data.
The analysis in the general case would require a model independent approach,
as in Ref.~\cite{Morgan:1993td}, to determine the number and location of poles
associated to the resonance.
However, previous analyses \cite{Morgan:1993td,Zou:1993az} have shown
that the $f_0(980)$ can be described in terms of a Breit-Wigner-like
resonance with two poles associated to it.
The Breit-Wigner parameterization is nevertheless a particular case of an
amplitude fulfilling unitarity.
Indeed, for a relativistic particle, the general solution to the unitarity
constraint can be written as \cite{Badalian:xj}
\begin{equation}
\label{amplab}
T_{ab}=\frac{e^{2i\delta_a(s)}-1}{2i\sqrt{\beta_a\beta_b}}\delta_{ab}-
       \frac{e^{i(\delta_a(s)+\delta_b(s))}}{\sqrt{\beta_a\beta_b}}
       \frac{\sqrt{G_a G_b}}{F(s)+iG(s)}\ ,
\end{equation}
where $\delta_{a,b}$ stands for the phase shifts describing the background in 
channels $(a,b)$, $\beta_{a,b}=\sqrt{1-4m_{a,b}^2/s}$ with $m_{a,b}$ the masses
involved in the two-body decays of the two channels, 
$F(s)$ and $G(s)$ are arbitrary real functions of $s$, and
$G_i(s)$ are positive functions with the property $G(s)=\sum_{i=a,b} G_i(s)$.
Identifying $F(s)=s-m_p^2$ and $G(s)=m_p\Gamma_p$, the amplitude (\ref{amplab})
reduces in the one channel case to the amplitude (\ref{apole})
up to an overall normalization factor.

The second of the ingredients mentioned above will be used in our framework
in order to identify the functions $F(s)$ and $G(s)$ to the real and imaginary 
parts of the complete one-loop propagator respectively.
The previous identification has the advantage of incorporating automatically 
thresholds effects (see below). This procedure requires the use of an effective field theory
in order to calculate the full propagator of the $f_0(980)$.
In general, effective field theories are of limited use in the description of
hadron physics where one expects the interactions to be strong.
There are cases, however, where these theories can be used.
The treatment of width effects, when the  width to mass ratio (taken as an expansion parameter) 
is small, is an example where effective field theories can be useful for the description of
narrow resonances, but not for broad ones. Notice in this respect that for the $f_0(980)$
$\Gamma_{f_{0}}/m_{f_{0}}\approx$ 0.04--0.1 \cite{Groom:2000in}.
Concerning the final form of the scalar propagator, 
the use of a simple Breit-Wigner parametrization with constant width,
which is applicable only to narrow resonances far from thresholds,
is not enough due to the closeness of the $K\bar K$ threshold and the $f_{0}$ mass.
Instead, one could use an energy dependent width, incorporating the kinematic dependences 
on the energy, but this approach amounts to include only the imaginary part of the
self-energy.
In our analysis, we prefer to use the fully corrected one-loop propagator,
including both the real and imaginary parts of the self-energy,
since this approach allows us to have a consistent description of the analytical
properties,
\textit{i.e.}~it provides a proper analytic continuation of the scattering amplitude
below the $K\bar K$ threshold.

After Dyson summation, the propagator of a scalar particle is \cite{Peskin:ev}
\begin{equation}
\label{sp}
\Delta(p^2)=\frac{i}{p^2-m_0^2+\Pi(p^2)}\ ,
\end{equation}
where $m_0$ is the \emph{bare} or tree-level mass of the resonance and
$\Pi(p^2)$ is the one-particle-irreducible (1PI) two-point function.
In the \emph{on-shell scheme}, a Taylor expansion of the real part of 
$\Pi(p^2)$ around the resonance mass allows to rewrite the scalar propagator
as
\begin{equation}
\label{spBW}
\Delta(p^2)=\frac{iZ}{p^2-m_R^2+im_R\Gamma_R}+\cdots\ ,
\end{equation}
where the renormalized mass $m_R$ (the so-called \emph{on-shell mass})
and the wave-function renormalization factor $Z$ are defined as
\begin{equation}
\begin{array}{l}
m_R^2=m_0^2-{\rm Re}\Pi(m_R^2)\ ,\\[1ex]
Z^{-1}=1+{\rm Re}\Pi^\prime(m_R^2)\ ,
\end{array}
\end{equation}
with ${\rm Re}\Pi^\prime(p^2)=d{\rm Re}\Pi(p^2)/dp^2$.
By analogy with a Breit-Wigner resonance, the width is defined
by\footnote{This definition applies only to narrow resonances,
$\Gamma_R\ll m_R$,
where ${\rm Im}\Pi(p^2)$ can be approximated by ${\rm Im}\Pi(m_R^2)$
over the width of the resonance.
If the resonance is broad, the full energy dependence of $\Pi(p^2)$ must be
taken into account.}
\begin{equation}
\label{GammaR}
\Gamma_R=\frac{1}{m_R}Z{\rm Im}\Pi(m_R^2)\ .
\end{equation}
However, this \emph{on-shell} definition of the resonance width is inadequate
since it vanishes when a two-particle $s$-wave threshold is approached from
below \cite{Bhattacharya:1993gr}. Due to the failure of the Taylor expansion of 
$\Pi(p^2)$ around $m_R^2$, Eq.~(\ref{GammaR}) has not the desired behaviour for a 
width properly defined. This is precisely the case under consideration
since the $K\bar K$ threshold lies in the vicinity of the $f_0(980)$ mass.

On the contrary, the \emph{pole approach} provides a consistent definition
of the resonance width that behaves sensibly in the threshold region.
In this approach, the Taylor expansion of $\Pi(p^2)$ is not performed and
the scalar propagator (\ref{sp}) is written as
\begin{equation}
\label{sppole}
\begin{array}{rcl}
\Delta(p^2)&=&\frac{i}{p^2-m_0^2+\Pi(p^2)}\\[1ex]
           &=&\frac{i}{p^2-m_R^2+{\rm Re}\Pi(p^2)-{\rm Re}\Pi(m_R^2)+
                       i{\rm Im}\Pi(p^2)}\ ,
\end{array}
\end{equation}
where $m_R^2=m_0^2-{\rm Re}\Pi(m_R^2)$.
Within the framework of the pole approach, the scattering amplitude (\ref{amplab}) 
describing  a resonance coupled to two channels is obtained by identifying the 
functions $F(s)$ and $G(s)$ with the denominator of the
complete one-loop scalar propagator in Eq.~(\ref{sppole}):
$F(s)=s-m_R^2+{\rm Re}\Pi(s)-{\rm Re}\Pi(m_R^2)$ and $G(s)={\rm Im}\Pi(s)$.
This procedure leads to
\begin{equation}
\begin{array}{rcl}
\label{amplabpole}
T_{ab}&=&\frac{e^{2i\delta_a(s)}-1}{2i\sqrt{\beta_a\beta_b}}\delta_{ab}-
         \frac{e^{i(\delta_a(s)+\delta_b(s))}}{\sqrt{\beta_a\beta_b}}
         \frac{\sqrt{s}\sqrt{\Gamma_a\Gamma_b}}
              {s-m_R^2+{\rm Re}\Pi(s)-{\rm Re}\Pi(m_R^2)+
              i{\rm Im}\Pi(s)}\\[1ex]
 &=&\frac{e^{2i\delta_a(s)}-1}{2i\sqrt{\beta_a\beta_b}}\delta_{ab}-
         \frac{e^{i(\delta_a(s)+\delta_b(s))}}{16\pi}
         \frac{g_a g_b}{s-m_R^2+{\rm Re}\Pi(s)-{\rm Re}\Pi(m_R^2)+
                        i{\rm Im}\Pi(s)}\ ,
\end{array}
\end{equation}
where $\Gamma_{a,b}=\frac{g_{a,b}^2}{16\pi\sqrt{s}}\beta_{a,b}$ are the 
partial decay widths of the resonance in channels $(a,b)$.
The common relation ${\rm Im}\Pi(p^2)=\sqrt{p^2}\Gamma(p^2)$
is not used to avoid confusion
(the width of the resonance in the pole approach is related to the
pole position, \emph{it is not} given by the tree level result following
from the optical theorem).
The renormalized mass $m_R$ and the tree-level coupling constants of the
resonance to the two channels $g_{a,b}$ are the parameters to be fitted
when confronting the scattering amplitude (\ref{amplabpole})
with data.
Once these parameters are extracted from experimental data,
the \emph{pole mass} $m_p$ and \emph{pole width} $\Gamma_p$ of the resonance
are obtained from the pole equation
\begin{equation}
\label{poleeq}
D(s_p)=s_p-m_R^2+{\rm Re}\Pi_+(s_p)-{\rm Re}\Pi_+(m_R^2)+
       i{\rm Im}\Pi_+(s_p)=0\ ,
\end{equation}
with $s_p=m_p^2-im_p\Gamma_p$ and $\Pi_+(s)\equiv \Pi(s+i\epsilon)$.
The pole equation (\ref{poleeq}) involves a complex function of a complex
variable. If for real $s$, $\Pi_+(s)=R(s)+iI(s)$,
then for arbitrary complex $s$
\begin{equation}
\label{Pi+}
\Pi_+(s)={\rm Re}R(s)-{\rm Im}I(s)+i\left[{\rm Im}R(s)+{\rm Re}I(s)\right]\ .
\end{equation}
In order to find \emph{all} the poles associated with a resonance coupled to
channels $(a, b)$ we have to look for the poles of Eq.~(\ref{poleeq}) in the
four different Riemann sheets defined by the complex channel momenta $p_{a,b}$.
Following the conventional classification, the sheets are enumerated according
to the signs of $({\rm Im}p_a,{\rm Im}p_b)$:
\begin{equation}
\begin{array}{lll}
{\rm sheet\ I}   & (+\,+): & ({\rm Im}p_a>0,{\rm Im}p_b>0)\ ,\\[1ex]
{\rm sheet\ II}  & (-\,+): & ({\rm Im}p_a<0,{\rm Im}p_b>0)\ ,\\[1ex]
{\rm sheet\ III} & (-\,-): & ({\rm Im}p_a<0,{\rm Im}p_b<0)\ ,\\[1ex]
{\rm sheet\ IV}  & (+\,-): & ({\rm Im}p_a>0,{\rm Im}p_b<0)\ .
\end{array}
\end{equation}
We restrict ourselves to the case of two-body channels involving particles of the
same mass.
Thus, the thresholds for channels $(a,b)$ are $2m_{a,b}$ and the momenta are 
defined as $p_{a,b}(s)=\sqrt{s-4m_{a,b}^2}/2$ ($m_a<m_b$ is assumed).
Since the complete propagator $D(s)$ is indeed an explicit function of the
momenta $p_{a,b}$, $D(s)=D[s,p_{a}(s),p_{b}(s)]$,
a change in the sign of the imaginary part of the momentum ---a change of sheet---
is achieved with the replacement of $p_{a,b}$ by $-p_{a,b}$ in the propagator.
Therefore, the poles are found solving the following four pole equations:
\begin{equation}
\begin{array}{ll}
D[s,p_{a}(s),p_{b}(s)]\ ,   & D[s,-p_{a}(s),p_{b}(s)]\ ,\\[1ex]
D[s,-p_{a}(s),-p_{b}(s)]\ , & D[s,p_{a}(s),-p_{b}(s)]\ .
\end{array}
\end{equation}

For the case of interest, namely the $f_0(980)$ scalar resonance coupled to 
a pair of pions and a pair of kaons\footnote{
In our analysis, we work in the isospin limit and therefore the mass difference
between $K^0$ and $K^+$ is not taken into account for the $K\bar K$ threshold.
The inclusion of this mass difference would deserve a more refined 
3-channel analysis with eight Riemann sheets that is beyond the scope of the
present work.},
the real and imaginary parts of the finite part of the 1PI two-point function
$\Pi(s)$ are
\begin{equation}
\label{RI}
\begin{array}{rcl}
R(s)&=&\frac{g_{f_0\pi\pi}^2}{16\pi^2}
\left[2-\beta_\pi\log\left(\frac{1+\beta_\pi}{1-\beta_\pi}\right)\right]+
      \frac{g_{f_0 K\bar K}^2}{16\pi^2}
\left[2-\beta_K\log\left(\frac{1+\beta_K}{1-\beta_K}\right)\right]
\Theta_K\\[1ex]
&+&\frac{g_{f_0 K\bar K}^2}{16\pi^2}
   \left[2-2\bar\beta_K\arctan\left(\frac{1}{\bar\beta_K}\right)\right]
   \bar\Theta_K\ ,\\[1ex]
I(s)&=&\frac{g_{f_0\pi\pi}^2}{16\pi}\beta_\pi+
       \frac{g_{f_0 K\bar K}^2}{16\pi}\beta_K\Theta_K\ ,
\end{array}
\end{equation}
where $\beta_i=\sqrt{1-4m_i^2/s}$ for $i=\pi, K$,
$\bar\beta_K=\sqrt{4m_K^2/s-1}$, $\Theta_K=\Theta(s-4m_K^2)$, and 
$\bar\Theta_K=\Theta(4m_K^2-s)$.
It is worth remarking that the step functions $\Theta$ are not introduced by
hand but result from the present calculation and play a crucial r\^ole in the
determination of the pole structure.

So far we have discussed the framework needed for the description of a
resonance coupled to two channels.
However, there are not current experiments that allow for
a direct comparison of two-particle scattering amplitudes with experimental data.
Therefore, in order to carry out the numerical analysis one has to rely on
production processes such as $J/\psi\rightarrow\phi(\pi\pi,K\bar K)$
and $J/\psi\rightarrow\omega (\pi\pi,K\bar K)$
or central production in proton-proton scattering $pp\rightarrow p(\pi\pi,K\bar K)p$.
As stated in the Introduction, we will perform our analysis using only the
former $J/\psi\rightarrow\phi(\pi\pi,K\bar K)$ decays as a mechanism for producing
pairs of pions and kaons.
In this respect, we follow Ref.~\cite{Morgan:1993td}
to relate the production amplitude $F$, also constrained by unitarity,
to the scattering amplitudes $T_{ab}$ in Eq.~(\ref{amplabpole}).
The corresponding amplitudes for $J/\psi\rightarrow\phi(\pi\pi,K\bar K)$
are then written as
\begin{equation}
\label{Fs}
\begin{array}{l}
F_{\pi}\equiv F(J/\psi\rightarrow\phi\pi^+\pi^-)=
\sqrt{\frac{2}{3}}\left[\alpha_{\pi}(s)T_{\pi\pi}+\alpha_{K}(s)T_{K\pi}\right]\ ,\\[1ex]
F_{K}\equiv F(J/\psi\rightarrow\phi K^+K^-)=
\sqrt{\frac{1}{2}}\left[\alpha_{\pi}(s)T_{\pi K}+\alpha_{K}(s)T_{KK}\right]\ ,
\end{array}
\end{equation}
where the real coupling functions $\alpha_{\pi,K}(s)$ are parametrized as
$\alpha_{i}(s)=\gamma_{i0}+\gamma_{i1}s$ and the $\gamma_{i}$ are obtained
from the fit.
Note that the theoretical expression for the $\pi\pi$ phase shift and inelasticity
are obtained from the $T_{\pi\pi}$ scattering amplitude in Eq.~(\ref{amplab}).
The $\pi\pi$ data used in our fits are extracted from three different experimental
analyses, two of them \cite{Hyams:zf,Grayer:at} based on data from the reaction
$\pi^-p\rightarrow \pi^+\pi^- n$ \cite{Grayer:1974cr} and the third one
\cite{Kaminski:1996da,Kaminski:2001hv} from
$\pi^-p_{\uparrow}\rightarrow \pi^+\pi^- n$ \cite{Becker:1978ks} and
$\pi^-p\rightarrow \pi^0\pi^0 n$ \cite{Gunter:2000am}.

\section{Numerical analysis}
\label{numerics}
Before proceeding with the numerical analysis we should keep in mind that our
method is based on the pole approach and thus, in order to obtain a background 
independent fit to the data, we need to restrain ourselves to a neighborhood 
of the $f_0(980)$ resonance.
For this reason we have chosen to work with experimental data on 
$J/\psi\rightarrow\phi\pi^+\pi^-$ and $J/\psi\rightarrow\phi K^+K^-$ decays
\cite{Falvard:1988fc,MK3}
and on the $\pi\pi$ phase shift and inelasticity
\cite{Hyams:zf,Grayer:at,Kaminski:1996da,Kaminski:2001hv}
in the range $0.8\leq \sqrt{s}\leq 1.1$ GeV.
It is worth noticing that within the kinematical region between 0.8 GeV and the
point where the rising of the $\pi\pi$ phase shift starts due to the appearance
of the $f_{0}(980)$, the contribution of the $\sigma$ scalar resonance to the
phase shift in this region can be reasonably described in terms of an energy
polynomial\footnote{
Moreover, the choice of using experimental data only around 980 MeV avoids the
need of describing the broad bump seen in the $\pi\pi$ phase shift around 600 MeV
for which a simple polynomial parametrization is not adequate.} (see below).

For the $\pi\pi$ phase shift we have used three different sets of data.
The first two sets differ mainly in the point lying just around 980 MeV and correspond
to solutions B \cite{Hyams:zf} (including this controversial point)
and D \cite{Grayer:at} (not including it) of Ref.~\cite{Grayer:1974cr}.
The last set of data corresponds to the ``down-flat'' solution of
Refs.~\cite{Kaminski:1996da,Kaminski:2001hv} that seems to be the most preferable
solution after a joint analysis of the $S$-wave $\pi^+\pi^-$ and $\pi^0\pi^0$ data
\cite{Becker:1978ks,Gunter:2000am}.
In the following we will denote these three sets of $\pi\pi$ phase shift data
as sets B, D and DF respectively.
Part of the differences concerning the pole parameters of the $f_0(980)$ resonance
reported in the literature could be due to the use of different data.
In this work, in order to quantify the influence of the $\pi\pi$ phase shift
used, we have performed fits using data-sets B, D and DF.

In the data fitting, a background term can be introduced for one or both channels
and furthermore different energy dependences of the phase shifts can be considered.
In our analysis we have included a background for the $\pi\pi$ and $K\bar K$
channels both with an energy dependence ranging from constant $(\delta=b_0)$
to quadratic $(\delta=b_0+b_1 s+b_2 s^2)$.
For the $\pi\pi$ case, the contribution of the $\sigma$ resonance to the background
term is included in this way.
For the $K\bar K$ case, only $b_1$ and $b_2$ are independent parameters
since the background term must vanish by continuity below the kaon threshold.
So then, the maximum number of parameters of our fits is 12:
the renormalized mass $m_R$,
the coupling constants $g_{f_0\pi\pi}$ and $g_{f_0 K\bar K}$,
the parameters $b^\pi_0, b^\pi_1$ and $b^\pi_2$ for the background term for pions,
$b^K_1$ and $b^K_2$ for the kaon background,
and the constants $\gamma_{\pi 0}, \gamma_{\pi 1}$ and $\gamma_{K 0}, \gamma_{K 1}$
parametrizing the $F_{\pi}$ and $F_{K}$ amplitudes.

\begin{table}
\centerline{
\begin{tabular}{|l|r|r|r|}\hline\hline
Fit                                 & set B            & set D            & set DF           \\ \hline
$m_R^2$ (GeV$^2$)                   & $0.966\pm 0.003$ & $0.982\pm 0.001$ & $0.978\pm 0.003$ \\ \hline
$g_{f_0\pi\pi}^2/16\pi$ (GeV$^2$)   & $0.071\pm 0.007$ & $0.065\pm 0.005$ & $0.11\pm 0.01$   \\ \hline
$g_{f_0 K\bar K}^2/16\pi$ (GeV$^2$) & $0.25 \pm 0.03$  & $0.16 \pm 0.01$  & $0.31\pm 0.04$   \\ \hline
$b_0^\pi$                           & $1.89 \pm 0.09$  & $1.47 \pm 0.03$  & $0.49\pm 0.05$   \\ \hline
$b_1^\pi$ (GeV$^{-2}$)              & $-1.1 \pm 0.2$   & $1.47 \pm 0.09$  & $4.6\pm 0.1$     \\ \hline
$b_2^\pi$ (GeV$^{-4}$)              & $0.8  \pm 0.1$   & $-0.50\pm 0.06$  & $-2.83\pm 0.09$  \\ \hline
$b_1^K$ (GeV$^{-2}$)                & $28.7 \pm 0.6$   & $15.5 \pm 0.3$   & $26.6\pm 0.4$    \\ \hline
$b_2^K$ (GeV$^{-4}$)                & $-12.6\pm 0.2$   & $-6.8 \pm 0.1$   & $-11.5\pm 0.2$   \\ \hline
$\gamma_{\pi 0}$                    & $5.1  \pm 0.4$   & $-0.2 \pm 0.9$   & $5.0\pm 1.0$     \\ \hline
$\gamma_{\pi 1}$ (GeV$^{-2}$)       & $-1.5 \pm 0.5$   & $8.2  \pm 1.0$   & $1.7\pm 1.0$     \\ \hline
$\gamma_{K 0}$                      & $-27.2\pm 0.5$   & $-39.5\pm 0.7$   & $-21.7\pm 0.5$   \\ \hline
$\gamma_{K 1}$ (GeV$^{-2}$)         & $29.7 \pm 0.5$   & $45.0 \pm 0.7$   & $26.2\pm 0.6$    \\ \hline
$\chi^2/$d.o.f                      & 1.10             & 0.87             & 1.10             \\ \hline
\hline
\end{tabular}
}
\caption{Values for the parameters
obtained from a joint fit to the $J/\psi\rightarrow\phi(\pi\pi,K\bar K)$ decays
and the $\pi\pi$ phase shift and inelasticity.
Set B, D or DF refers to the set of $\pi\pi$ phase shift data used in the fit.}
\label{tablefit}
\end{table}

\begin{figure}
\centerline{\includegraphics[width=\textwidth]{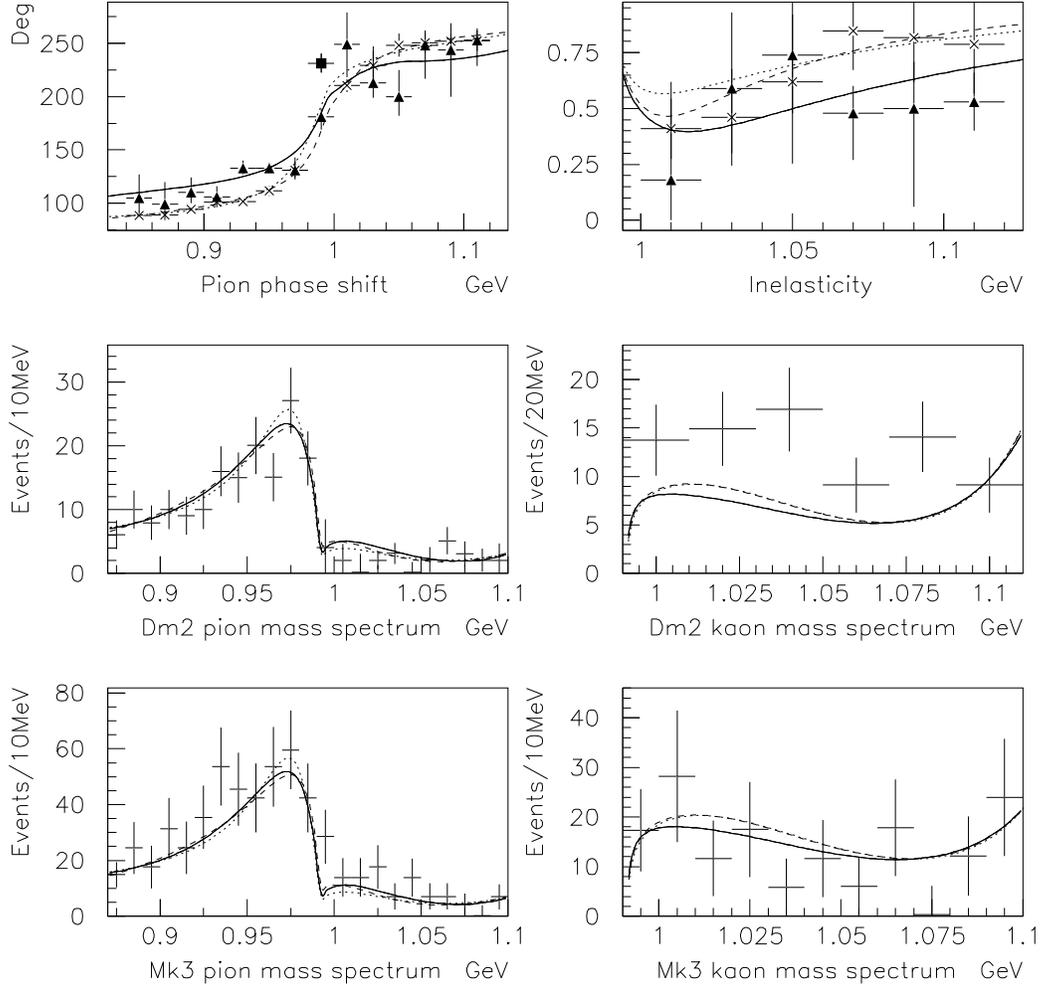}}
\caption{Fit to $\pi\pi$ data on the phase shift and inelasticity
(cross \protect\cite{Grayer:at}, solid square \protect\cite{Hyams:zf},
solid triangle \protect\cite{Kaminski:1996da,Kaminski:2001hv})
and to data on $J/\psi\rightarrow\phi(\pi\pi,K\bar K)$ production
\protect\cite{Falvard:1988fc,MK3}.
Fits to the data sets B, D and DF are shown with dotted, dot-dashed and solid lines
respectively.}
\label{figfit}
\end{figure}

The results of the different fits performed show that
\emph{i)} the $\chi^2$ improves when a $K\bar K$ background is included,
although changing its energy dependence makes no relevant difference;
\emph{ii)} the values obtained for the physically relevant parameters
($m_{R}$, $g_{f_0\pi\pi}$ and $g_{f_0 K\bar K}$) change only a few percent
when different backgrounds are considered.
The outcome of the fit for the case of quadratic backgrounds for pions and kaons
is written in Table \ref{tablefit} and shown in Fig.~\ref{figfit}.
The values for the renormalized mass $m_R$,
the coupling constants $g_{f_0\pi\pi}$ and $g_{f_0 K\bar K}$,
the $\chi^2$ per degree of freedom and the other fitted parameters
are presented for sets B, D and DF of $\pi\pi$ data.
Concerning the values of the coupling constants obtained from the fit,
we observe that the coupling of the $f_0(980)$ to kaons is stronger than the coupling to
pions: $g_{f_0 K\bar K}^2/g_{f_0\pi\pi}^2=3.52, 2.46, 2.82$ for sets B, D and DF respectively.
Model dependent values for these coupling constants have been recently reported by the
SND \cite{Achasov:2000ym}, CMD-2 \cite{Akhmetshin:1999di} and KLOE \cite{Aloisio:2002bt}
Collaborations.
Their analyses, based on the study of the $\phi\rightarrow\pi^0\pi^0\gamma$
radiative process, give
$g_{f_0 K^+K^-}^2/g_{f_0\pi^+\pi^-}^2=4.6\pm 0.8$ (SND),
$3.61\pm 0.62$ (CMD-2) and $4.00\pm 0.14$ (KLOE)\footnote{The coupling constants
$g_{f_0\pi\pi(K\bar K)}$ used in our analysis are related to the more common
coupling constants $g_{f_0\pi^+\pi^-(K^+K^-)}$ and $g_{\pi(K)}$ used in
experimental analyses by
$\frac{2}{3}\,\frac{g_{f_0\pi\pi}^2}{4\pi}=\frac{g_{f_0\pi^+\pi^-}^2}{4\pi}=
\frac{4}{3}\,g_{\pi}\,m_{f_{0}}^2$ and
$\frac{1}{2}\,\frac{g_{f_0 K\bar K}^2}{4\pi}=\frac{g_{f_0 K^+K^-}^2}{4\pi}=
g_{K}\,m_{f_{0}}^2$.}.
Other analyses, based either on $pp\rightarrow p(\pi\pi,K\bar K)p$ central production
\cite{Barberis:1999cq} or on $f_0(980)$ production in hadronic $Z_0$ decay
\cite{Ackerstaff:1998ue}, suggest the same behaviour for the coupling constants
and obtain $g_{K}/g_{\pi}=2.1\pm 0.6$ and $g_{K}/g_{\pi}\simeq 10$ respectively.
On the contrary, the analysis by the E791 Collaboration \cite{Aitala:2000xt}
on the $f_0(980)$ production in $D_{s}\rightarrow 3\pi$ decays gives
$g_{K}/g_{\pi}=0.2\pm 0.6$.
The values we obtain for both coupling constants and for their ratio are smaller than
model predictions \cite{Tornqvist:1999tn,Napsuciale:1998ip,Shabalin:ey}
and also than previous determinations \cite{Kaminski:1998ns}.

Let us now proceed to determine the mass and width of the $f_0$ resonance within the
pole approach.
Fits to the data sets B, D and DF lead to values for the renormalized mass of 
$m_R=983\pm 2$ MeV, $m_R=991\pm 1$ MeV and $m_R=989\pm 2$ MeV respectively.
The pole parameters of the resonance are determined once the values of the renormalized mass
and the ones regarding the coupling constants (see Table \ref{tablefit})
are included in the pole equation (\ref{poleeq}).
The numerical solution of the pole equation yields for data sets B, D and DF:
\begin{equation}
\label{poles}
\begin{array}{ll}
m_p^B=987\pm 3\ \mbox{MeV}\ , & \Gamma_p^B=42\pm 9\ \mbox{MeV}\ ,\\[1ex]
m_p^D=999\pm 2\ \mbox{MeV}\ , & \Gamma_p^D=39\pm 8\ \mbox{MeV}\ ,\\[1ex]
m_p^{DF}=1001\pm 6\ \mbox{MeV}\ , & \Gamma_p^{DF}=52\pm 16\ \mbox{MeV}\ .
\end{array}
\end{equation}
This determination of the pole mass and width of the $f_0(980)$ resonance together with
its couplings constants to the $\pi\pi$ and $K\bar K$ channels constitute
the main result of this work.
Our results in Eq.~(\ref{poles}) are in fair agreement with several values of the $f_0(980)$
pole parameters appeared recently in the literature.
The values $m_p=994\ \mbox{MeV}$ and $\Gamma_p=28\ \mbox{MeV}$ are obtained from an analysis of
meson-meson interactions in a nonperturbative chiral approach \cite{Oller:1998hw}.
Similar analyses give $m_p=987\ \mbox{MeV}$ and $\Gamma_p=28\ \mbox{MeV}$ \cite{Oller:1998zr} or
$m_p=981.4\ \mbox{MeV}$ and $\Gamma_p=44.8\ \mbox{MeV}$ \cite{Oller:1997ti}.
Other analysis based on the study of meson-meson interactions in different coupled channel unitarity
models give $m_p=1015\pm 15\ \mbox{MeV}$ and $\Gamma_p=86\pm 16\ \mbox{MeV}$ \cite{Anisovich:1997zw},
$m_p=991\pm 3\ \mbox{MeV}$ and $\Gamma_p=71\pm 14\ \mbox{MeV}$ \cite{Kaminski:1998ns},
$m_p=1008\ \mbox{MeV}$ and $\Gamma_p=54\ \mbox{MeV}$ \cite{Locher:1997gr},
$m_p=993.2\pm 6.5\pm 6.9\ \mbox{MeV}$ and $\Gamma_p\sim 100\ \mbox{MeV}$ \cite{Ishida:1995xx} or
$m_p=1006\ \mbox{MeV}$ and $\Gamma_p=34\ \mbox{MeV}$ \cite{Tornqvist:1995ay}.

\begin{figure}
\centerline{\includegraphics[width=0.9\textwidth]{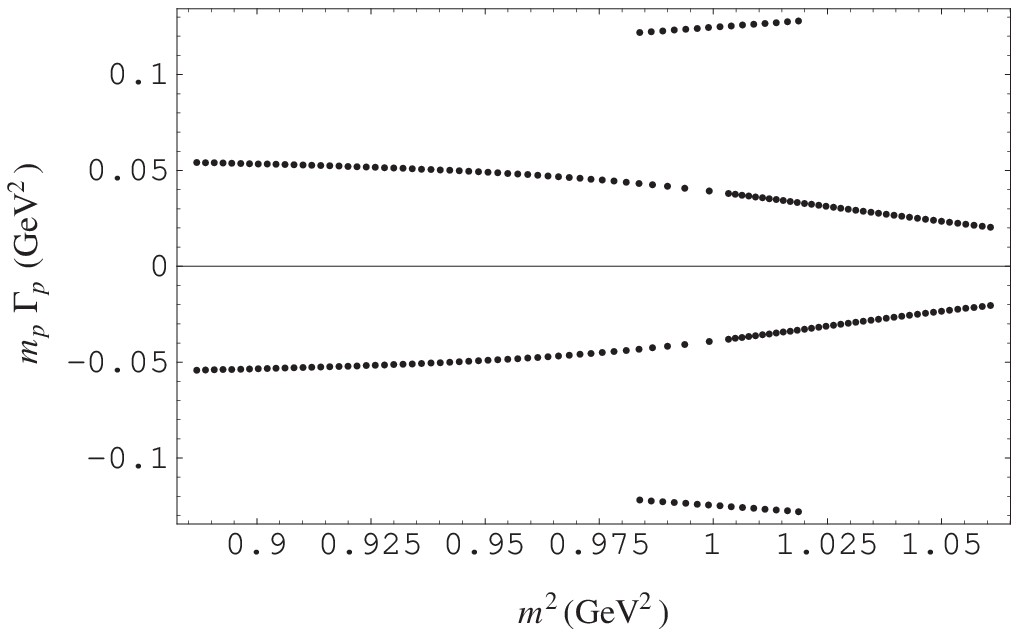}}
\caption{Behavior of the pole position ($m_p^2$ \textit{versus} $m_p\Gamma_p$) 
as a function of the renormalized mass $m_R$ in the range $941\leq m_R\leq 1027$ MeV.
The set of points with larger values of $|m_p\Gamma_p|$ appearing above 0.975 GeV$^2$
correspond to values of $m_R>1020$ MeV and are associated to poles in sheet III.}
\label{figpole}
\end{figure}

In addition to this result we have also analyzed the variation of the pole position as a function
of the renormalized mass $m_R$,
\textit{i.e.}~the coupling constants $g_{f_0\pi\pi}$ and $g_{f_0 K\bar K}$ are kept fixed to their
values for set D in Table \ref{tablefit}.
In Fig.~\ref{figpole}, we show $m_p^2$ \textit{versus} $m_p\Gamma_p$,
which are related to the real and imaginary parts of the pole $s_p$,
for values of the renormalized mass in the range $941\leq m_R\leq 1027$ MeV.
Thus, each point on the plot corresponds to a solution of the pole equation (\ref{poleeq})
---in terms of the values obtained for $m_p$ and $\Gamma_p$--- for a given value of $m_R$.
Only the physically relevant pole is shown in Fig.~\ref{figpole};
complex conjugate poles or any other kind of poles are not included.
In order to generate Fig.~\ref{figpole}, we looked for solutions of the pole equation in each
of the four Riemann sheets reaching the following conclusions:
\begin{enumerate}
\item
We did not find a pole neither in sheet I nor IV in the vicinity of 980 MeV
(we looked for poles in the range 960--1020 MeV).
\item
We find poles in sheet II in the range $941\leq m_R\leq 1027$ MeV.
From these solutions we see that the pole mass is always larger than the renormalized mass,
\textit{i.e.}~$m_p^{II}>m_R$.
When $4m_\pi^2<m_R^2<4m_K^2$, we obtain a pole width $\Gamma_p$ that has to be identified with the
$f_0\rightarrow\pi\pi$ decay width.
The width so found does not coincide with the decay width calculated from the tree level expression
$\Gamma_{f_0\rightarrow\pi\pi}=g_{f_0\pi\pi}^2\beta_\pi/16\pi\sqrt{s}$,
the difference arising from the contribution of the Im$R(s)$ term in Eq.~(\ref{Pi+})
once $s$ takes a complex value.
\item
We find poles in sheet III \emph{only} for $m_R>1020$ MeV.
In this case, the pole mass is always of the order of 20 MeV smaller than the renormalized mass,
\textit{i.e.}~$m_p^{III}<m_R$, and $\Gamma_p$ should be identified with the tree level width
$\Gamma_{f_0\rightarrow\pi\pi}+\Gamma_{f_0\rightarrow K\bar K}$.
Again, this decay width does not coincide with the pole width for the same reasons as before.
\end{enumerate}
From points 2 and 3 above we conclude that only one pole in sheet II will be necessary to describe
the $f_0(980)$ resonance, since poles in sheet III only appear for $m_R>1020$ MeV while our fit to
data always yields $m_R<1$ GeV.
Moreover, these points also indicate that the \emph{overlapping of poles is not possible}.
The idea of the overlapping of poles states that values of the renormalized mass $m_R$ close but
below the kaon threshold $(m_R<2m_K)$ may lead to values of the pole mass $m_p$ above the threshold
$(m_p>2m_K)$ and that values of $m_R$ close but above the threshold $(m_R>2m_K)$ may lead to values
of $m_p$ below the threshold $(m_p<2m_K)$.
We stress that our conclusion \emph{is not} general,
it depends upon the values used for the coupling constants, in particular on the ratio
$g_{f_0 K\bar K}^2/g_{f_0\pi\pi}^2$ \cite{eden&taylor}.

\section{Conclusions}
Using a generalized version of the Breit-Wigner parametrization based upon unitarity and
a propagator obtained from effective field theory including loop contributions,
we have performed a fit to experimental data on the $\pi\pi$ phase shift, the inelasticity and
the $J/\psi\rightarrow\phi\pi\pi$ and $J/\psi\rightarrow\phi K\bar K$ decays.
The fit has been restricted to a neighborhood of 980 MeV ($0.8\leq \sqrt{s}\leq 1.1$ GeV)
thus providing a process and background independent way of extracting the intrinsic
properties (pole mass and width) of the $f_{0}(980)$ scalar resonance.

The solution of the pole equation for the values of the parameters resulting from the fit
allows us to conclude that the $f_{0}(980)$ is described in terms of a \emph{single pole in sheet II}
and yields the values
$m_{f_{0}}^{B}=987\pm 3$ MeV and $\Gamma_{f_{0}}^{B}=42\pm 9$ MeV,
$m_{f_{0}}^{D}=999\pm 2$ MeV and $\Gamma_{f_{0}}^{D}=39\pm 8$ MeV, and
$m_{f_{0}}^{DF}=1001\pm 6$ MeV and $\Gamma_{f_{0}}^{DF}=52\pm 16$ MeV
for the data-sets B, D and DF respectively.
We also analyzed the behaviour of the pole position as a function of the renormalized mass
$m_{R}$ and found that a pole in sheet III only arises when $m_{R}>1020$ MeV
while our fit to data yields always values $m_R<1$ GeV.

\section*{Acknowledgements} 
R.~E.~acknowledges A.~Farilla, F.~Nguyen, G.~Venanzoni and N. Wu for valuable discussions,
A.~Bramon for a careful reading of the manuscript, L.~L\'esniak for providing us with the latest
$\pi\pi$ data, and J.~R.~Pel{\'a}ez for the Mathematica program for extracting the poles
from the scattering amplitudes. 
Work partly supported by the EU, EURODAPHNE (TMR-CT98-0169) and EURIDICE (HPRN-CT-2002-00311) networks. 
Work also partly supported by the Ministerio de Ciencia y Tecnolog\'{\i}a and FEDER (FPA2002-00748),
and CONACYT.

\end{document}